\documentclass[11pt]{article}
\usepackage{graphicx}
\usepackage{latexsym}
\usepackage{amsfonts}

\newcommand{\qed}{\hfill $\Box$}

\newtheorem{theorem}{Theorem}[section]
\newtheorem{lemma}[theorem]{Lemma}
\newtheorem{proposition}[theorem]{Proposition}
\newtheorem{corollary}[theorem]{Corollary}

\newtheorem{assumption}[theorem]{Assumption}

\newtheorem{remark}[theorem]{Remark}

\begin{document}

\title{Smooth Value Functions for a Class of Nonsmooth
Utility Maximization Problems}
\author{Baojun Bian\thanks{Department of Mathematics,
         Tongji University, Shanghai 200092, China.
bianbj@tongji.edu.cn, Research of this author was supported by
NSFC No.10671144 and National Basic Research Program of China
(2007CB814903)}, Sheng Miao   and Harry Zheng\thanks{Department of
Mathematics, Imperial College, London SW7 2BZ, UK.
s.miao@imperial.ac.uk and h.zheng@imperial.ac.uk.}}
\date{}
\maketitle

\noindent{\bf Abstract.}
In this paper we prove that there exists a
  smooth classical solution to the HJB equation
  for a large class of constrained problems with utility
functions that are not necessarily differentiable or strictly concave.
The value function is  smooth if admissible controls satisfy an integrability condition  
or if it is continuous on the closure of its domain.
The key idea is  to work on the dual control problem and
the dual HJB equation. We construct a smooth, strictly convex solution
to the dual HJB equation and show that its conjugate function is a smooth,
strictly concave solution to the primal HJB equation  satisfying the
terminal and boundary  conditions.

\medskip\noindent{\bf Key words.}
nonsmooth utility maximization, classical solution to HJB equation,
smooth value function, dual control problem.

\medskip\noindent{\bf AMS subject classifications.} 90C46, 49L20

\section{Introduction}
There has been extensive research in utility maximization. Two main methods are stochastic control and convex duality. The stochastic control approach requires the underlying state process be Markovian and applies the dynamic programming principle and Ito's lemma to derive a nonlinear parabolic PDE (HJB equation) for the optimal value function. If there is a classical  solution to the HJB equation one may then apply the verification theorem to show that the value function is smooth and find the optimal control as a byproduct. The convex duality approach requires the objective functional be concave and the state process be linear. It first solves a static maximization problem and applies convex analysis to show the existence of the optimal solutions to the primal and dual problems and establishes their dual relationship. It then uses the martingale representation theorem or more general
optional decomposition theorem to super-replicate the optimal terminal wealth/consumption. For excellent expositions of these two methods in utility maximization,
see \cite{fs,ks98,ks, pham} and references therein.

The smoothness of the value function is a highly desirable property. One normally has to impose some conditions to ensure that. One key condition is the uniform ellipticity of the diffusion coefficient, which is not
 satisfied for the standard wealth process as long as doing nothing is a feasible portfolio trading strategy.
When the trading constraint set is a closed convex cone
and the utility function  is strictly concave and continuously differentiable and satisfies some growth conditions and the market is complete the value function is a smooth solution to the HJB equation, see \cite{ks98}. When the constraint set is the whole space and
the utility function is of  power or logarithmic type, the value function has a  closed-form expression. The approach in \cite{ks98} crucially depends on the differentiability and strict concavity of the utility function
as the inverse function of the marginal utility is extensively used.

For general non-smooth and/or non-strictly-concave utility functions  it is not clear if there exist smooth solutions to the HJB equation.
To deal with the lack of a priori knowledge of the differentiability of the value function one may use a weak solution concept and  characterize the value function as a unique viscosity solution to the HJB equation. Due to the remarkable stability property of the viscosity solution one may solve the HJB equation numerically. It is in general difficult to show the differentiability of the value function even it is known to be a viscosity solution to the HJB equation (but see the remarkable paper \cite{ss}). The lack of the differentiability of the value function makes impossible to apply the verification theorem to find the optimal control.

Consider a financial market consisting of one bank account  and $n$ stocks.
The discounted price process  $S=(S^1,\ldots,S^n)'$ of $n$ risky assets
is  modelled by
$$
dS_t={\rm diag}(S_t)(b(t) dt+ \sigma(t) dW_t),\quad 0\leq t\leq T
$$
with the initial price $S_0=s$,
where ${\rm diag}(S_t)$ is a diagonal $n\times n$ matrix with diagonal elements $S^i_t$,
$b$ and $\sigma$ are deterministic continuous
vector and nonsingular matrix valued functions of time $t$,
representing the stock excess returns and volatilities, respectively,  and
 $W$ is  a $n$-dimensional standard Brownian motion
 on a complete probability space
$(\Omega,\mathcal{F},P)$, endowed with a natural filtration
$\{\mathcal{F}_t\}$ generated by $W$.
The  discounted wealth process $X$  satisfies the SDE
\begin{equation} \label{wealth}
dX_t = X_t(\pi_t' b(t) dt + \pi_t'\sigma(t) dW_t),
\quad X_0=x_0
\label{wealth}
\end{equation}
where $\pi_t=(\pi^1_t,\ldots,\pi^n_t)'$ are progressively measurable
control processes satisfying    $\pi_t\in K$, a closed convex
    cone, a.s. for $t\in[0,T]$.

A standard utility maximization problem is given by
\begin{equation} \label{primal}
\sup E[U(X_T)]
\mbox{ subject to (\ref{wealth})}
\end{equation}
where $U$ is a utility function which
is continuous, increasing, concave, and $U(0)=0$.

Denote by $V(t,x)$ the value function of
(\ref{primal}) for $0\leq t\leq T$ and $x\geq0$.
The corresponding HJB equation is given by
\begin{equation} \label{HJB}
 -V_t - H(t,x,V_x(t,x),V_{xx}(t,x)) =0,\ x>0,t<T
\end{equation}
with the terminal condition $V(T,x)=U(x)$ and the boundary condition
$V(t,0)=0$, where $H$ is  a Hamiltonian, defined by
\begin{equation} \label{hamiltonian}
H(t,x,p,M)=\sup_{\pi\in K}
\{\pi'b(t)xp + {1\over 2}|\sigma(t)'\pi|^2x^2 M\}
\end{equation}
and
$V_t$ is a partial derivative of $V$ with respect to $t$, $V_x$ and $V_{xx}$ are defined similarly.

The contribution of this paper is that we show that there exists a
 smooth classical solution to the HJB equation
 (\ref{HJB})  for a large class of constrained problems with utility
functions that are not necessarily differentiable or strictly
concave (Theorem~\ref{maintheorem}).  The value function is  smooth if admissible controls satisfy an integrability condition  (Theorem~\ref{verification})
or if it is continuous on the closure of its domain
(Theorem~\ref{smoothV}).
The key idea is  to work on the dual control problem and
the dual HJB equation. We show that
there is a smooth, strictly convex solution
to the dual HJB equation and its conjugate function is a smooth,
strictly concave solution to the primal HJB equation  satisfying the
terminal and boundary  conditions.
 We use an observation which asserts that under certain
structure conditions the solution of parabolic partial
differential equations is smooth and strictly convex even if the
initial date is not differentiable or strictly convex. This is
related to the convexity preserving and constant rank principle
for solution of partial differential equations, see \cite{BG08,
BG09, JT04, LM}

The rest of the paper is organized as follows.  Section 2 reviews the
existence results of nonsmooth utility maximization and characterizes the dual control problem. Section 3 constructs the smooth solutions to the primal and dual HJB equations. Section 4 proves the verification theorem
under an integrability condition for admissible controls. Section 5 shows
that the primal and dual value functions are smooth if they are continuous on the closure of the domains with a comparison method. Section 6 gives two applications, one in the efficient frontier of utility and conditional value-at-risk, and the other in the monotonicity of absolute risk aversion measures.

\section{Dual Control Problem}
In this section we briefly review the main results on the existence of the optimal solutions to the primal and dual problems and characterize the dual control problem.
We focus on the dual domain for the application of stochastic control theory.
Almost all work in literature on utility maximization are for continuously differentiable and strictly concave utility functions. The main references for nonsmooth utility maximization are \cite{btz, dpt, wz09, wz10}.

To use the duality method to study the value function of the utility maximization
problem (\ref{primal}) we need first to formulate a dual minimization problem
with a well defined dual domain. The choice of the dual domain is often problem
specific. For a complete market generated by Brownian motions \cite{ks98}
chooses dual variables as
$$ H_{t,\nu} = \exp\left(-\int_0^t \theta'_{s,\nu}dW_s -{1\over 2} \int_0^t
|\theta_{s,\nu}|^2 ds - \int_0^t\xi(\nu_s)ds\right)
$$
where $\theta_{t,\nu}=\sigma^{-1}_t(b_t+\nu_t)$, $\xi(v)=\sup_{p\in K}(-p'v)$
is the support function of $-K$, and $\nu$ are $\mathcal{F}_t$ progressively
measurable processes satisfying $E[\int_0^T \xi(v_t)dt]<\infty$.
This gives a natural set of dual variables.  The
approach in \cite{ks98}
crucially depends on the assumption that
the utility function $U$ is differentiable and strictly
concave and some other conditions.
Results of \cite{ks98} cannot be directly applied to the problem
of this paper. On the other hand \cite{ks98} provides an explicit construction
of the dual process which turns out to be very useful in proving the dual
relation of the primal and dual value functions.

For general semimartingale asset price processes
 the duality method is normally used to show the existence of the optimal solutions to the primal and dual problems and to establish  their dual
relation. There are several definitions of dual variables depending on the
primal problem formulation. \cite{ks} chooses the set of dual variables consisting
of nonnegative supermartingale
processes $Y$ with $Y_0=y$ such that $XY$ are supermartingales for  all admissible wealth processes $X$ with initial endowment $x$, while \cite{btz} takes
nonnegative random variables $Y$ in $L^1$ such that $E[X_TY]\leq xy$ for
all admissible terminal wealth $X_T$.

Consider a security market consisting of $d+1$ assets, one bond and $d$ stocks. Assume bond price $S^0$ equals one  and (discounted) stock price
$S=(S^i)_{1\leq i\leq d}$ is modeled by a $(0,\infty)^d$ valued
semimartingale on a filtered probability space
$(\Omega,\mathcal{F},(\mathcal{F}_t)_{0\leq t\leq T},\mathbb{P})$.
Let $K$ be a closed convex  cone. Denote by $\Theta$
 the set of admissible trading strategies such that
 every $\theta \in \Theta$ is a predictable process, integrable with
respect to $S$ and valued in  $K$ a.s.~for all $t$.

The wealth process is defined by initial capital x and admissible strategy $\theta$ as follows
$$X_t^{x,\theta}=x+\int_0^t \theta_udS_u.$$
The set of nonnegative wealth processes with initial value $x$ is defined by
\begin{equation} \label{primaldomain}
\mathcal{X}_+(x):=\{X^{x,\theta}:\theta\in\Theta \mbox{ and }
X_t^{x,\theta}\geq 0\mbox{ for all }t\in[0,T]\}
\end{equation}
and the set of terminal values of nonnegative wealth processes is defined by
$\mathcal{X}_+^T(x):=\{X_T^{x,\theta}:X^{x,\theta}\in \mathcal{X}_+(x)\}$.
The problem of maximizing the expected utility of the terminal wealth
is given by
$$ V(x)=\sup_{X\in \mathcal{X}_+^T(x)} E[U(X)]$$
where $U$ is an increasing concave utility function
defined on the positive real line.
The dual problem is formulated as
$$
\tilde V(y):=\inf_{Y\in\mathcal{Y}_+^T(y)} E[\tilde{U}(Y)]
$$
where $\tilde U$ is the dual function of $U$, defined by
$$\tilde{U}(y):=\sup_{x\geq 0}\{U(x)-xy\}$$
and $\mathcal{Y}_+^T(y)$ is the set of dual variables, defined by
$$
\mathcal{Y}_+^T(y):=\{Y\in L_+^0: E[XY]\leq xy \mbox{ for all } x\in \mathbb{R}_+
\mbox{ and }X\in \mathcal{X}_+^T(x)\}.
$$

We now state the main theorem on the existence and the dual
relation of the primal and dual problems.
\begin{theorem}
\label{auxthm}
Assume some technical conditions are satisfied (see \cite{btz},
Theorem 3.2 ). Then
\begin{enumerate}
\item There exist $\bar y\geq0$ and $\bar Y\in\mathcal{Y}_+^T(\bar y)$ such
that $\tilde V(\bar y)=E[\tilde{U}(\bar Y)]$ and
$W(x)=\tilde V(\bar y) + x\bar y$, where
$W(x):=\inf_{y>0} (\tilde V(y) +xy)$.
\item There exists $\bar X\in\mathcal{X}^T_+(x)$  such that
$V(x)=E[U(\bar X)]$.
\item  $V(x)=W(x)$, $E[\bar X\bar Y]=x\bar y$, and
$\bar X\in -\partial\tilde{U}(\bar Y)$.
\end{enumerate}
\end{theorem}

\begin{remark}{\rm The technical conditions are related to
the no-arbitrage of a financial market, the existence of
a constrained optional decomposition, and the asymptotic elasticity of
utility functions, see
\cite{btz}, Theorem 3.2 and  \cite{wz09}, Theorem 5.1, for details.
}\end{remark}

\begin{proposition}\label{pr:g0}
Assume the same technical conditions
and that $U$ is strictly increasing.
If $\bar Y$ is an optimal dual solution of $W(x)$, then $\bar Y>0$ a.s.
\end{proposition}
\noindent{\it Proof.} We first show that if $\bar Y(\omega)=0$, for
some $\omega\in \Omega$, the optimal $\bar X(\omega)=\infty$. From
\cite{wz09}, Theorem 5.1,
$\eta=\bar X(\omega)\in-\partial\tilde{U}(\bar Y(\omega))=-\partial\tilde{U}(0)$.
$\tilde{U}$ is a convex function, then
$\tilde{U}(z)\geq\tilde{U}(0)-\eta(z-0)$, $\forall z>0$.
$\tilde{U}(z)\geq U(\infty)-\eta z$.
$\tilde{U}(z)=\sup_{x>0}(U(x)-xz)=U(\bar{x})-\bar{x}z$ if and only if
$z\in\partial U(\bar{x})$. Assume $\eta<\infty$, choose any
$\bar{x}>\eta$, then choose $\bar{z}\in\partial U(\bar{x})$,
$\bar{z}>0$. We have $\tilde{U}(\bar{z})=U(\bar{x})-\bar{x}\bar{z}\geq
\tilde U(0)-\eta \bar{z}$, which implies
$0>U(\bar{x})-U(\infty)\geq
\bar{z}(\bar{x}-\eta)>0$. This is a contradiction. We can now show
that $\bar Y>0$ a.s. Assume $\exists A\subset\Omega$,
$\mathbb{P}(A)>0$, $\bar Y(\omega)=0,\omega\in A$, then
$\bar X(\omega)=\infty,\omega\in A$ from the discussion above.
For any equivalent probability measure $Q$, we
have $E_Q[\bar X]= E_Q[\bar X1_A]+E_Q[\bar X1_{A^c}]=\infty$, a contradiction to
the budget constraint and the no-arbitrage condition. \qed

The dual domain $\mathcal{Y}_+^T(y)$ is a set of random variables. To formulate a dual control, we need to have the dual domain consisting of stochastic processes, not just random variables. It is suggested in \cite{ks}
that a natural dual process domain is
\begin{equation} \label{dualdomain}
\mathcal{Y}_+(y)=
\{Y\geq0:\,\,Y_0=y\,\mbox{ and }\,XY
 \mbox{ is a supermartingale, for all }X\in\mathcal{X}_+(x)\}.
\end{equation}
This indeed serves our purpose. We have
the following equivalent results of Theorem \ref{auxthm}.

\begin{theorem} \label{dual}
Assume the same technical conditions are satisfied. Then
\begin{enumerate}
\item There exist $y^*\geq0$ and $Y^*\in\mathcal{Y}_+(y^*)$ such
that $\tilde V(y^*)=E[\tilde{U}(Y_T^*)]$
and $W(x)=\tilde V(y^*) +xy^*$, where
$W(x):=\inf_{y>0} (\tilde V(y) +xy)$.
\item There exists $X^*\in\mathcal{X}_+(x)$  such that
$V(x)=E[U(X_T^*)]$.
\item  $V(x)=W(x)$, $E[X_T^*Y_T^*]=xy^*$, and
$X_T^*\in -\partial\tilde{U}(Y_T^*)$.
\end{enumerate}
\end{theorem}

\noindent{\it Proof.} Define
$$\hat V(y)=\inf_{Y\in \mathcal{Y}_+(y)}E[\tilde{U}(Y_T)].$$ It is obvious that if $Y\in \mathcal{Y}_+(y)$ then $Y_T\in \mathcal{Y}_+^T(y)$ and we have $\tilde V(y)\leq \hat V(y)$.
From Theorem \ref{auxthm} (1), there exist $\bar y\geq 0$
and $\bar Y\in\mathcal{Y}_+^T(\bar y)$ such
that $W(x)=E[\tilde{U}(\bar Y)+x\bar y]$.
Since
$E[X\bar Y]\leq x\bar y$ for all $x>0$ and $X\in\mathcal{X}_+^T(x)$ we have
$$\bar Y\in \{h\in L_+^0(\Omega,\mathcal{F},P):0\leq h\leq Y_T,\mbox{ for some } Y\in \mathcal{Y}_+(\bar y)\}
$$
by \cite{ks}, Proposition 3.1 (ii). Let $y^*=\bar y$.
We can find a $Y^*\in \mathcal{Y}_+(y^*)$ such that $\bar Y\leq \tilde Y_T^*$. Since $\tilde U$ is a decreasing function we have
$E[\tilde{U}(\bar Y)]\geq
E[\tilde{U}(Y_T^*)]$, which implies that
$\tilde V(y)\geq \hat V(y)$. Therefore
$$ \hat V(y) \leq E[\tilde{U}(Y_T^*)]
\leq E[\tilde{U}(\bar Y)] = \tilde V(y) \leq \hat V(y).$$
That gives $\hat V(y)=\tilde V(y)= E[\tilde{U}(Y_T^*)]
=E[\tilde{U}(\bar Y)]$ and (1) is proved.
(3) can be proved in the same way as that of \cite{wz09}, Lemma 5.8 and
is  omitted here.  \qed

\begin{remark}{\rm We know that $\bar Y\leq Y_T^*$ but
we have not claimed that $\bar Y=Y_T^*$. This would be the case if one chose a naive and seemly natural stochastic process $Y_t^*
=E[\bar Y|\mathcal{F}_t]$ for $0\leq t\leq T$. However,
it is not clear if $X^*Y^*$ is a supermartingale with this construction. }\end{remark}

We now continue to use the control process $\pi_t$ instead of
$\theta_t$ which are related by
$\pi^i_t=\frac{\theta^i_tS^i_t}{X_t}$. The domains of the prime and
dual problems are given by (\ref{primaldomain}) and (\ref{dualdomain}), respectively.
Since the filtration is generated by diffusion processes the
Doob-Mayer decomposition theorem
implies that the positive supermartingale $Y\in\mathcal{Y}_+(y)$
can be decomposed as:
$$
Y_t=y\varepsilon(-\alpha' W)_tD_t
$$
where $\varepsilon$ is the Dol\'eans-Dade  exponential  and
$dD_t=-\beta_t D_tdt$, $\beta_t\geq 0$, and $D_0=1$, i.e., $Y$ satisfies the SDE
$$dY_t=Y_t(-\alpha_t' dW_t-\beta_tdt).$$

\begin{proposition}
Let $\tilde{K}$ be the positive polar cone of $K$,
i.e., $\tilde{K} =\{p: p' v \geq 0,\;\forall v\in K\}$.
Then the optimal value of the dual problem can be characterized by
$$ \tilde V(y)=\inf_{Y\in \bar{\mathcal{Y}}_+(y)}E[\tilde{U}(Y_T)]
$$
where $\bar{\mathcal{Y}}_+(y)$ is the set of processes $Y$ satisfying
$$
dY_t=-Y_t(\sigma(t)^{-1}v_t+\theta(t))' dW_t, \quad Y_0=y
$$
and $v$ are progressively measurable with $v_t\in\tilde{K}$ a.s. for all $t$,
and $\theta(t)=\sigma(t)^{-1}b(t)$.
\end{proposition}
\noindent{\it Proof.} Obviously, $\tilde V(y)=
\inf_{Y\in\mathcal{Y}_+(y)}E[\tilde{U}(Y_T)]$.
When $X\in\mathcal{X}_+(x)$ and $Y\in\mathcal{Y}_+(y)$, Ito's lemma implies
$$
dX_tY_t=
X_tY_t(\pi_t' b(t)-\pi_t'\sigma(t) \alpha_t-\beta_t)dt+X_tY_t(\pi_t'\sigma(t)-\alpha_t')dW_t.
$$
Since  $XY$ is a supermartingale and nonnegative, we must have
$\pi_t' b(t)-\pi_t'\sigma(t) \alpha_t-\beta_t\leq 0$ a.s. for all $t$.
Define $v_t:=-\sigma(t)(\theta(t)-\alpha_t)$. Then $\alpha_t=\sigma(t)^{-1}v_t+\theta(t)$. Supermartingale property of $XY$ gives
$$
-\pi_t'v_t\leq \beta_t\mbox{,      }\,\,\forall\,\pi_t\in K\mbox{   }\,\, a.s.
$$
$\beta_t$ is the upper bound of $-\pi_t'v_t$ for all $\pi_t\in K$.
Since $K$ is a cone we must have $v_t\in\tilde{K}$. Otherwise,
if on a positive measure set $A\in\mathcal{F}$, $v_t$ is not in $\tilde{K}$. Then the support function $\zeta(v_t)=\sup_{p_t\in K}(-\pi_t'v_t)=\infty$, which implies $\beta_t=\infty$ on  $A$, a contradiction.
Define
$$\mathcal{D}:=\{D\geq 0,\,dD_t=-\beta_tD_tdt,\,\beta_t\geq 0,\,D_0=1\}$$ then $Y=\bar{Y}D\in\mathcal{Y}_+(y)$, $\bar{Y}\in\bar{\mathcal{Y}}_+(y)$ and $D\in\mathcal{D}$.
$$
\tilde V(y)=\inf_{Y\in\mathcal{Y}_+(y)}E[\tilde{U}(Y_T)]
=\inf_{\bar{Y}\in\bar{\mathcal{Y}}_+(y),\,D\in\mathcal{D}}
E[\tilde{U}(Y_TD_T)]
\leq \inf_{\bar{Y}\in\bar{\mathcal{Y}}_+(y),\,D=1}
E[\tilde{U}(\bar{Y}_T)].
$$
On the other hand, $\tilde{U}$ is decreasing and $D_t\leq 1$. Thus
$$
\inf_{Y\in\mathcal{Y}_+(y)}E[\tilde{U}(Y_T)]
\geq\inf_{Y\in\bar{\mathcal{Y}}_+(y)}E[\tilde{U}(Y_T)].
$$
So $\tilde V(y)=\inf_{Y\in\bar{\mathcal{Y}}_+(y)}E[\tilde{U}(Y_T)]$. \qed

Denote by $\tilde V(t,y)$ the value function of the dual problem, i.e.,
$$\tilde V(t,y)=\inf_{Y\in \bar{\mathcal{Y}}_+(y)}E[\tilde{U}(Y_T)|Y_t=y].$$
Then the dual HJB equation is given by
\begin{equation} \label{tHJB}
\tilde{V}_t+ \inf_{\tilde{\pi}\in
\tilde{K}}\{\frac{1}{2}|\theta(t)+\sigma(t)^{-1}\tilde{\pi}|^2y^2\tilde{V}_{yy}\}=0,\
y>0,\; t<T
\end{equation}
with the terminal condition $\tilde{V}(T,y)= \tilde{U}(y)$.
It is easy to verify that $\tilde V(t,y)$ is convex in $y$ for fixed $t\in
[0,T]$. Denote by $\hat{\pi}(t)$ the unique
minimizer of $f(\tilde{\pi})=|\theta(t)+\sigma(t)^{-1}\tilde{\pi}|^2$ over $\tilde{\pi}\in \tilde{K}$ and
$\hat{\theta}(t)=\theta(t)+\sigma(t)^{-1}\hat{\pi}(t)$. The  equation
(\ref{tHJB}) is then equivalent to a linear PDE
\begin{equation} \label{dualHJB}
\hat{V}_t+\frac{1}{2}|\hat{\theta}(t)|^2 y^2\hat{V}_{yy}=0,\
y>0,\; 0\leq t<T.
\end{equation}

\section{Smooth Solutions to  HJB Equation}
We assume  that $U$ and $\hat\theta$ satisfy the following conditions.
\begin{assumption} \label{utility}
Utility function
$U$ is a continuous, increasing, concave function on $[0,\infty)$
satisfying $U(0)=0$, $U(\infty)=\lim_{x\rightarrow \infty}U(x)=\infty$, and
\begin{equation}\label{growth}
0\leq U(x)\leq L(1+x^p),\ x\geq 0
\end{equation}
for some constants $L>0$, $0<p<1$.
\end{assumption}

\begin{assumption}
\label{parabolicity} $\hat{\theta}$
is continuous on $[0,T]$  and there is a positive
constant $\theta_0$ such that $|\hat{\theta}(t)|\geq \theta_0$
for all $t\in [0,T]$.
\end{assumption}

\begin{remark}\label{rk3.2}
{\rm
Condition $U(0)=0$ can be replaced
by  $U(0)>-\infty$. If $U\in C^1(0,\infty)$, then $U'(\infty)=0$
from (\ref{growth}). We do not assume that the Inada condition holds.
If $U$ satisfies Assumption \ref{utility}, then
$\tilde{U}$ is a continuous decreasing convex function satisfying
$\tilde{U}(0)=\infty$, $\tilde{U}(\infty)=0$, and
\begin{equation} \label{dualgrowth}
0\leq \tilde{U}(y)\leq \sup_{x> 0}\{ L(1+x^p)-yx\}
\leq \hat{L} (1+y^{{p\over p-1}}),\ y>0
\end{equation}
where $\hat{L}=\max\{L,(Lp)^{\frac{1}{1-p}}(p^{-1}-1)\}$.
 In particular, if  $\tilde{U}\in C^1$, then $\tilde{U}'(\infty)=0$.
}\end{remark}

Consider the linear SDE
\begin{equation} \label{dualsde}
d\hat{Y}_s= -\hat{Y}_s\hat{\theta}(s)' d W_s, \quad s\geq t
\end{equation}
with the initial value $\hat{Y}_t=y$. Denote by
$\hat Y^{t,y}_s$ the unique strong solution to  (\ref{dualsde}) and define
a function $\hat V$ on $[0,T]\times (0,\infty)$ by
$$
\hat{V}(t,y)=E[\tilde{U}(\hat{Y}^{t,y}_T)].
$$

\begin{lemma} \label{hatVbound}
$\hat V$ satisfies
\[
0 \leq \hat{V}(t,y)\leq K (1+y^{\frac{p}{p-1}}),\ t\in[0, T]
\]
for some positive constant $K$. Furthermore, $\hat V$ is continuous on $[0,T]\times (0,\infty)$ and is a viscosity solution to  the linear PDE (\ref{dualHJB}).
\end{lemma}
\noindent {\it Proof.} Define
$Z_s=\left(\hat Y^{t,y}_s\right)^{{p\over p-1}}$ for $s\geq t$. Ito's lemma implies that $Z$ satisfies the SDE
$$ dZ_s=Z_s\left({p\over 2(p-1)^2}|\hat\theta(s)|^2 ds
- {p\over p-1}\hat\theta(s)'dW_s\right)$$
with initial value $Z_t=y^{{p\over p-1}}$.
Therefore
$$Z_T=y^{{p\over p-1}}
\exp\left({p\over 2(p-1)^2}\int_t^T |\hat\theta(s)|^2 ds\right)H_t
$$
where
$$ H_t=\exp\left(\int_t^T \hat\eta(s)'dW_s -{1\over 2}\int_t^T
|\hat\eta(s)|^2 ds \right)
$$
and $\hat\eta(s)={p\over 1-p}\hat\theta(s)$. Since
$E(H_t)=E(H_T)=1$ we have from (\ref{dualgrowth}) that
$$\hat{V}(t,y)\leq
\hat{L} (1+E[Z_T])
\leq K (1+y^{\frac{p}{p-1}})
$$
where $K=\hat{L}e^{{p\over 2(p-1)^2}\int_0^T |\hat\theta(s)|^2 ds}$.
Furthermore,
for all $0\leq t\leq T$ and $y\geq y_0$ for any fixed $y_0>0$ we have
$$ E[(\tilde U(\hat Y^{t,y}_T))^2]
\leq K^2  E[(1+y^{\frac{p}{p-1}}H_t)^2]
\leq 2K^2 (1+ y_0^{\frac{2p}{p-1}} e^{\int_0^T |\hat\eta(s)|^2ds})
$$
which implies that $\{\tilde U(\hat Y^{t,y}_T):
0\leq t\leq T, \; y\geq y_0\}$ is a class of uniformly integrable random variables. From the continuity of $\tilde U$ and
$\hat Y^{t,y}_T$ with respect to $t$ and $y$ we conclude that
$\hat V$ is continuous on $[0,T]\times (0,\infty)$.
Since $\hat V(t,y)=E[V(\tau,\hat Y^{t,y}_\tau)]$ for any stopping time $\tau\geq t$ it is straightforward to show that $\hat V$ is a viscosity solution to (\ref{dualHJB}), see, for example, \cite{pham}.
\qed

Next we show that $\hat{V}$ is smooth and strictly
convex in $y$ and is a classical solution to  (\ref{dualHJB}).
 Since $\tilde{U}$ is only continuous and convex,
we must improve the regularity and convexity.
The regularity is well known in the PDE theory. The
key idea to improve convexity is connected to the convexity
preserving and constant rank principle for solutions of PDEs, see
\cite{BG08, BG09, JT04, LM}. The techniques used here are likely
to be useful in solving other problems involving nonlinear
equations.

\begin{lemma} \label{LemhatV}
The  function $\hat{V}$ is a classical solution to
(\ref{dualHJB}). Furthermore, $\hat{V}\in
C^{1,\infty}([0, T)\times (0,\infty))$ and satisfies
\[
\hat{V}_{y}(t,y)<0,\  \hat{V}_{yy}(t,y)>0
\]
for every $t\in[0, T)$.
\end{lemma}
\noindent {\it Proof.}
Define $v(t,z)=\hat{V}(t,e^z)$. Then
$v(t,z)\leq K (1+e^{\frac{p}{p-1}z})$ and $v$ is a continuous
viscosity solution to the linear  PDE
\[
v_t+\frac{1}{2}|\hat{\theta}(t)|^2 (v_{zz}-v_z)=0,\ z\in R,\;0\leq
t<T
\]
with the terminal condition $v(T,z)= \tilde{U}(e^z)$ for $z\in R$.
Let $\tau=\frac{1}{2}\int_t^T |\hat{\theta}(\eta)|^2 d\eta $ and
$\tilde{v}(\tau,z)=e^{\frac{\tau}{4}-\frac{z}{2}}v(t,z)$. Then the
equation for $v$ is reduced to the standard Cauchy problem
\[
\tilde{v}_\tau- \tilde{v}_{zz}=0,\ z\in R,\; t>0
\]
with the initial value
$\tilde{v}(0,z)=e^{-\frac{z}{2}}\tilde{U}(e^z)$. It is easy to see
that $\tilde{v}$ is a classical solution and $\tilde v\in
C^{\infty,\infty}((0,\infty)\times R)$.  Since $\tilde{v}(\tau,z)$
grows exponentially in $z$, we obtain from the Poisson formula
(\cite{Fri}, Chapter 1) that
\[
\tilde{v}(\tau,z)=\frac{1}{2\sqrt{\pi\tau}} \int_{-\infty}^\infty
e^{-\frac{(z-\xi)^2}{4\tau}} e^{-\frac{\xi}{2}}
\tilde{U}(e^\xi)d\xi.
\]
Hence
\[
v(t,z)=\frac{1}{2\sqrt{\pi\tau}} e^{-\frac{1}{4}\tau}
\int_{-\infty}^\infty e^{-\frac{(z-\xi)^2}{4\tau}}
e^{-\frac{1}{2}(\xi-z)} \tilde{U}(e^\xi)d\xi
\]
and $v\in C^{1,\infty}([0,T)\times R)$.
Finally we get
\begin{equation} \label{hatV}
\hat{V}(t,y)=\frac{1}{2\sqrt{\pi\tau}} e^{-\frac{1}{4}\tau}
 \int_0^\infty
e^{-\frac{(\ln \xi)^2}{4\tau}} \xi^{-\frac{3}{2}}
\tilde{U}(y\xi)d\xi
\end{equation}
and $\hat V\in C^{1,\infty}([0,T)\times (0,\infty))$.
Since $\tilde{U}$ is decreasing and convex, it follows that
$\hat{V}$ is decreasing and convex for fixed $t\in[0,T)$ from
(\ref{hatV}). Hence
\[
\hat{V}_{y}(t,y)\leq 0,\  \hat{V}_{yy}(t,y)\geq 0
\]
for every $t\in[0, T)$.
Differentiating (\ref{dualHJB}) twice, we conclude that
$w(t,y)=\hat{V}_{yy}(t,y)$ is a nonnegative classical solution to
the  equation
\[
w_t+\frac{1}{2}|\hat{\theta}(t)|^2 y^2w_{yy}+
2|\hat{\theta}(t)|^2yw_y +|\hat{\theta}(t)|^2 w
=0,\ y>0,\;0\leq t<T.
\]
If $w(t_0,y_0)=0$ for some $(t_0,y_0)$ with $t_0<T$, then
$(t_0,y_0)$ is a minimum point of $w$ and
$\hat{V}_{yy}(t,y)=w(t,y)=0$ for all $(t,y)\in (t_0, T)\times (0,\infty)$
by the strong maximum principle (\cite{Fri}, Chapter 2). This
implies that $\hat{V}(t,y)$ is linear in $y$ for any fixed $t\in
(t_0, T]$, in particular, $\tilde{U}(y)$ is linear. This is a
contradiction and we conclude that $\hat{V}_{yy}(t,y)>0$ for every
$t\in[0, T)$. Similarly, we deduce that $\hat{V}_{y}(t,y)<0$ for
every $t\in[0, T)$. \qed

\begin{lemma} \label{Vlimit} We have
\begin{equation} \label{hatVlimit}
\lim_{y\rightarrow 0} \hat{V}(t,y)=\infty, \lim_{y\rightarrow
\infty}\hat{V}(t,y)=0
\end{equation}
 and
$$
\lim_{y\rightarrow 0}\hat{V}_y(t,y)=-\infty, \lim_{y\rightarrow
\infty}\hat{V}_{y}(t,y)=0
$$
for  $t\in[0, T)$.
\end{lemma}

\noindent {\it Proof.} We have from (\ref{hatV}) and Remark~\ref{rk3.2}
that
\begin{eqnarray*}
\hat{V}(t,y)&\geq& \frac{1}{2\sqrt{\pi\tau}} e^{-\frac{1}{4}\tau}
 \int_0^1
e^{-\frac{(\ln \xi)^2}{4\tau}} \xi^{-\frac{3}{2}}
\tilde{U}(y\xi)d\xi\\
&\geq& \left(\frac{1}{2\sqrt{\pi\tau}} e^{-\frac{1}{4}\tau}
 \int_0^1
e^{-\frac{(\ln \xi)^2}{4\tau}} \xi^{-\frac{3}{2}} d\xi
\right)\tilde{U}(y)
\end{eqnarray*}
This implies that $\lim_{y\rightarrow 0} \hat{V}(t,y)=\infty$.

To prove $\lim_{y\rightarrow \infty}\hat{V}(t,y)=0$, we can
estimate, for $y>1$ and $a>0$, that
\begin{eqnarray*}
\int_0^a e^{-\frac{(\ln \xi)^2}{4\tau}}
\xi^{-\frac{3}{2}}\tilde{U}(y\xi)d\xi &\leq&  \hat{L} \int_0^a
e^{-\frac{(\ln \xi)^2}{4\tau}}\xi^{-\frac{3}{2}}
(1+y^{\frac{p}{p-1}}\xi^{\frac{p}{p-1}})d\xi\\
&\leq&  \hat{L} \int_0^a e^{-\frac{(\ln\xi)^2}{4\tau}}
\xi^{-\frac{3}{2}}(1+\xi^{\frac{p}{p-1}})d\xi
\end{eqnarray*}
 and
$$
\int_a^\infty e^{-\frac{(\ln \xi)^2}{4\tau}} \xi^{-\frac{3}{2}}
\tilde{U}(y\xi)d\xi \leq \left(\int_a^\infty e^{-\frac{(\ln
\xi)^2}{4\tau}} \xi^{-\frac{3}{2}} d\xi\right) \tilde{U}(ay).
$$
Combining these estimates, we conclude that $\lim_{y\rightarrow
\infty}\hat{V}(t,y)=0$.

Since $\hat{V}(t,y)$ is a convex smooth function in $y$ for fixed
$t\in [0,T)$, we conclude that $\hat{V}_y(t,y)$ is increasing in
$y$. Suppose $\lim_{y\rightarrow 0}\hat{V}_y(t,y)=A>-\infty$. Then
$\hat{V}_y(t,y)\geq A$ and $\hat{V}(1,t)-\hat{V}(t,y)\geq
\int_y^1\hat{V}_y(t,x)dx\geq A(1-y)$ for $0<y<1$. This contradicts
to (\ref{hatVlimit}). Similarly, we deduce that
$\lim_{y\rightarrow\infty}\hat{V}_{y}(t,y)=0$ for every $t\in[0,
T)$. \qed

Let $Y(t,\cdot)$ be the inverse function of $-\hat{V}_y(t,\cdot)$,
i.e.,
$$
-\hat{V}_y(t,Y(t,x))= x,\ Y(t,-\hat{V}_y(t,y))=y,
$$
for fixed $t\in[0, T)$. $Y(t,x)$ is well defined on
$[0,T)\times(0,\infty) $ from Lemmas \ref{LemhatV} and \ref{Vlimit}.
Since $\hat{V}\in C^{1,\infty}([0, T)\times
(0,\infty))$ and $\hat{V}_{yy}(t,y)>0$, the inverse function $Y\in
C^{1,\infty}([0, T)\times (0,\infty))$ by  the implicit
function theorem. Let
\begin{equation} \label{u}
u(t,x)= \inf_{y> 0}\{\hat{V}(t,y)+xy\}.
\end{equation}
We now show that $u$ is a classical solution to the HJB equation
(\ref{HJB}). We need the following result which is similar to
\cite{XS2}, Lemma 3.2.
\begin{lemma} \label{L1} Let  $a$ be a given number.
Let $\hat{\pi}(t)$ be the unique
minimizer of convex function

\[
f(\tilde{\pi})=|{\rm sgn}(a)\theta(t)+\sigma(t)^{-1}\tilde{\pi}|^2
=|\sigma(t)^{-1}({\rm sgn}(a)b(t)+\tilde{\pi})|^2
\]
over $\tilde{\pi}\in \tilde{K}$, where ${\rm sgn}(a)$
 is a sign function which equals 1 if $a>0$ and $-1$ if $a<0$.
Denote $\hat{\theta}(t)={\rm sgn}(a)\theta(t)+\sigma(t)^{-1}\hat{\pi}(t)$.
Then ${\pi}^*(t)=\frac{|a|}{2} Df(\hat{\pi}(t))=|a|
(\sigma(t)')^{-1}\hat{\theta}(t)$ is the
unique minimizer of convex function
\[
g(\pi)=\frac{1}{2}|\pi'\sigma(t)|^2-a\pi' b(t)
\]
over $\pi\in K$. Furthermore,
\[
g(\pi^*(t))=-\frac{1}{2}a^2 |\hat{\theta}(t)|^2.
\]
\end{lemma}
\noindent {\it Proof.} Since $\tilde{K}$ is a convex cone, we see
that $f(\eta\hat{\pi})$ attains its minimum at $\eta=1$. Hence $
\hat{\pi}(t)' Df(\hat{\pi}(t))=0$. Furthermore, for any given $q\in
\tilde{K}$, $f(\eta\hat{\pi}(t)+(1-\eta)q)$ attains its minimum at
$\eta=1$, which implies $q' Df(\hat{\pi}(t))\geq
\hat{\pi}(t)' Df(\hat{\pi}(t))=0$, we conclude that $Df(\hat{\pi}(t))\in K$.

Direct computation yields
\[
Df(\hat{\pi}(t))=2(\sigma(t)\sigma(t)')^{-1}({\rm sgn}(a)b(t)+\hat{\pi}(t))
=2(\sigma(t)')^{-1}\hat{\theta}(t)
\]
and
\[
Dg(\pi)= \sigma(t)\sigma(t)'\pi -ab(t).
\]
Let $\pi^*(t)=\frac{|a|}{2}Df(\hat{\pi}(t))
=|a|(\sigma(t)')^{-1}\hat{\theta}(t)$. Then
$\pi^*(t)\in K$ and simple algebra shows that
\[
(\pi^*(t))' Dg(\pi^*(t))=0,\ \pi' Dg(\pi^*(t))\geq 0
\]
for all $\pi\in K$, which implies that
 $\pi^*(t)$ is the unique minimizer of $g$ over
$\pi\in K$. Furthermore,
\[
g(\pi^*(t))=\frac{1}{2}a^2
|\hat{\theta}(t)|^2-a^2\hat{\theta}(t)'\sigma(t)^{-1}b(t)=-\frac{1}{2}a^2
|\hat{\theta}(t)|^2.
\]
\qed

We now state the main result of this section.
\begin{theorem}
\label{maintheorem}
 Assume $K$ is a closed convex cone and
Assumptions \ref{utility} and \ref{parabolicity} hold. Then there
exists a function  $u\in C^0([0,T]\times
[0,\infty)) \cap C^{1,2}([0,T)\times (0,\infty))$ which is a
classical solution to the HJB equation (\ref{HJB}). The maximum of
the Hamiltonian $H$ is achieved at
$$\pi^*(t,x)=-(\sigma(t)')^{-1}\hat\theta(t){u_x(t,x)\over x u_{xx}(t,x)}$$ and $\pi^*(t,x)\in K$. Furthermore,
$u(t,x)$ is strictly increasing and strictly concave in $x$ for
fixed $t\in [0,T)$ with $u(T,x)=U(x)$ and $u(t,0)=0$, and $0\leq
u(t,x)\leq \tilde{K}(1+x^p)$ for some  constant $\tilde{K}$.
\end{theorem}

\noindent {\it Proof.} Let $u(t,x)$ be defined by (\ref{u}).
We have, for $(t,x)\in[0,T)\times(0,\infty)$,
$$
u(t,x)=\hat{V}(t,Y(t,x))+xY(t,x),
$$
which yields the regularity of $u(t,x)$. Direct computation yields
$$
u_t(t,x)=\hat{V}_t(t,Y(t,x)),\quad
u_{x}(t,x)=Y(t,x), \quad
u_{xx}(t,x)=-\frac{1}{\hat{V}_{yy}(t,Y(t,x))}.
$$
Since $Y(t,x)>0$ and $\hat{V}_{yy}(t,Y(t,x))<0$ for fixed $0\leq
t<T$, the function $u(t,\cdot)$ is strictly increasing and
strictly concave. Substituting $y=Y(t,x)$ into equation
(\ref{dualHJB}) we get
$$
u_t-\frac{1}{2}|\hat{\theta}(t)|^2\frac{u_x^2}{u_{xx}}=0.
$$
We conclude by lemma \ref{L1} that $u$ is a classical solution to the
HJB equation (\ref{HJB}) and the maximum of the Hamiltonian is achieved at $\pi^*(t,x)$.  Furthermore, from Lemma \ref{hatVbound}.
\[
u(t,x)\leq \inf_{y> 0}\{K(1+y^{\frac{p}{p-1}})+xy\}
\leq \tilde{K}(1+x^p)
\]
where $\tilde{K}=K+{1\over p}\left({1-p\over Kp}\right)^{p-1}$.
\qed

\section{Verification Theorem}
Theorem \ref{maintheorem} confirms that there is a classical
solution $u$ to the HJB equation and the Hamiltonian achieves its
maximum at a point $\pi^*$ in $K$, i.e., there is a classical
solution to the nonlinear PDE
$$ u_t-{1\over 2}|\hat\theta(t)|^2{u_x^2\over u_{xx}}=0,\quad
x>0,t<T.$$
We now show that the value function $V$ is indeed a smooth classical solution
to the HJB equation (\ref{HJB}) with the optimal feedback control $\pi^*$.
Since 
the drift and diffusion terms in SDE (\ref{wealth}) do not satisfy the uniform Lipschitz continuous and linear growth conditions due to the unboundedness of the control set $K$, we do not know if solutions to SDE (\ref{wealth}) are square integrable and cannot directly apply the method of localization and the dominated convergence theorem to prove the verification theorem, see \cite{pham} for details. We assume that the following additional condition
be satisfied for admissible trading strategies $\pi$:
\begin{equation} \label{novikov}
 E\left[\exp\left({1\over 2}\int_0^T |\pi_t'\sigma(t)|^2 dt\right)\right]< \infty.
\end{equation}
Condition (\ref{novikov}) is stronger than the usual square integrability condition $E[\int_t^T|\pi_t'\sigma(t)|^2dt]<\infty$. 
It can be shown that the set of all admissible controls $\pi$ satisfying
(\ref{novikov}) is a convex set.  
 We can now state the verification theorem.
\begin{theorem} \label{verification}
Let $u$ be given as in Theorem \ref{maintheorem} and admissible controls satisfy (\ref{novikov}). Then $V(t,x)\leq u(t,x)$ on $[0,T]\times (0,\infty)$. Furthermore, if $\pi^*$ satisfies (\ref{novikov}) and
SDE (\ref{wealth})
admits a unique nonnegative strong solution with the feedback control $\pi^*$.
Then $V(t,x)=u(t,x)$ on $[0,T]\times [0,\infty)$ and $\pi^*$ is an optimal Markovian control.
\end{theorem}

\noindent{\it Proof}. Since $u$ is a smooth
classical solution to the HJB equation (\ref{HJB})
 we have for all $(t,x)\in [0,T)\times (0,\infty)$ and $\pi\in K$, that
\begin{equation} \label{hjb}
u_t(t,x)+u_x(t,x)x\pi'b(t)+ {1\over 2}u_{xx}(t,x)x^2|\pi'\sigma(t)|^2\leq
0
\end{equation}
and the equality  holds in (\ref{hjb}) if  $\pi=\pi^*(t,x)$. For any
 $s\in [t,T)$, stopping time $\tau \in [t,\infty)$, and admissible control 
 $\pi$ satisfying
 (\ref{novikov}), we have, by Ito's lemma
and (\ref{hjb}), that
\begin{equation} \label{veri}
 u(s\wedge \tau, X^{t,x}_{s\wedge \tau})
\leq u(t,x) 
+ \int_t^{s\wedge\tau} u_x(v,X^{t,x}_v)X^{t,x}_v\pi_v'\sigma(v)dW_v 
\end{equation}
where $X^{t,x}_{s}$ is the solution of SDE (\ref{wealth}) with the trading
strategy $\pi$ and the initial condition $X^{t,x}_t=x$. Let 
$$\tau=\tau_n=\inf\{s\geq t: \int_t^s 
| u_x(v,X^{t,x}_v)X^{t,x}_v\pi_v'\sigma(v)|^2 dv
\geq n\},$$
then the stopped process 
$\{\int_t^{s\wedge\tau} u_x(v,X^{t,x}_v)X^{t,x}_v\pi_v'\sigma(v)dW_v,
t\leq s\leq T\}$ 
 is a martingale. Taking expectation in (\ref{veri}) leads to
\begin{equation} \label{uwedge}
 E[u(s\wedge \tau_n, X^{t,x}_{s\wedge \tau_n})]
\leq u(t,x).
\end{equation}
Since $0<p<1$ we may choose $\alpha\in (1,1/p)$ and show, by Theorem \ref{maintheorem}
and the convexity of function $x^\alpha$, that
\begin{eqnarray*}
u(s\wedge \tau_n, X^{t,x}_{s\wedge \tau_n})^\alpha
&\leq& (\tilde K(1+ (X^{t,x}_{s\wedge \tau_n})^p))^\alpha\\
&=& \tilde K^\alpha 2^{\alpha-1} 
(1+ (X^{t,x}_{s\wedge \tau_n})^{\alpha p})\\
&=& \tilde K^\alpha 2^{\alpha-1} 
\left(1+ x^{\alpha p} H^\pi_{s\wedge \tau_n}\exp(\int_t^{s\wedge \tau_n}
 \Psi_v dv)\right)
\end{eqnarray*}
where 
\begin{eqnarray*}
H^\pi_{s\wedge \tau_n} &=& 
\exp\left(\int_t^{s\wedge \tau_n}(\alpha p)\pi_v'\sigma(v)dW_v 
-{1\over 2}(\alpha p)^2|\pi_v'\sigma(v)|^2 dv \right)\\
\Psi_v &=& (\alpha p)\pi_v' b(v) - {1\over 2}(\alpha p)
|\pi_v' \sigma(v)|^2 + {1\over 2}(\alpha p)^2|\pi_v'\sigma(v)|^2.
\end{eqnarray*}
Simple algebra shows that
\begin{eqnarray*}
\Psi_v 
&=& - {1\over 2}(\alpha p)(1-\alpha p)
|\sigma(v)'\pi_v -{1\over 1-\alpha p} \sigma(v)^{-1}b(v)|^2
+{\alpha p\over 2(1-\alpha p)}|\sigma(v)^{-1}b(v)|^2\\
&\leq& {\alpha p\over 2(1-\alpha p)}|\theta(v)|^2.
\end{eqnarray*}
Therefore,
$$ u(s\wedge \tau_n, X^{t,x}_{s\wedge \tau_n})^\alpha
\leq \tilde K^\alpha 2^{\alpha-1} 
\left(1+ x^{\alpha p} H^\pi_{s\wedge \tau_n}
\exp(\int_t^T {\alpha p\over 2(1-\alpha p)}|\theta(v)|^2 dv)\right).
$$
Finally, since $\pi$ satisfies (\ref{novikov}) and $0<\alpha p<1$ we know that $H^\pi_{s\wedge \tau_n}$ 
is a martingale from Novikov's condition, which implies 
$$ E[u(s\wedge \tau_n, X^{t,x}_{s\wedge \tau_n})^\alpha]
\leq \tilde K^\alpha 2^{\alpha-1} 
\left(1+ x^{\alpha p} 
\exp(\int_t^T {\alpha p\over 2(1-\alpha p)}|\theta(v)|^2 dv)\right).
$$
We conclude that $\{u(s\wedge \tau_n, X^{t,x}_{s\wedge \tau_n}): n\geq 1\}$
is a family of uniformly integrable random variables. Since
$\tau_n\uparrow \infty$ a.s. as  $n\to\infty$ and
$u\in C^0([0,T]\times [0,\infty)$ we may let $n$ tend to infinity in
(\ref{uwedge}) to get 
$$ E[u(s,X^{t,x}_s)]\leq u(t,x).$$
We can  apply exactly the same discussion as above and let $s$ tend to $T$, also note the terminal condition, to get
$$ E[U(X^{t,x}_T)]\leq u(t,x).$$
From the arbitrariness of admissible control $\pi$ we deduce that 
$V(t,x)\leq u(t,x)$. 

Next denote by $\bar X^{t,x}_s$ the solution to SDE (\ref{wealth}) 
with the trading strategy $\pi^*$. From Ito's lemma, 
(\ref{hjb}) with the equality, and the same discussion as above we have
$$ E[u(t,\bar X^{t,x}_s)]
= u(t,x).$$
Letting $s$ tend to $T$ we get
$$ u(t,x)=E[U(\bar X^{t,x}_T)] \leq V(t,x).$$
We have proved that $V(t,x)=u(t,x)$ and the optimal feedback control is $\pi^*(t,x)$. \qed

\section{Smoothness of  Value Functions}
In this section we  show that if the value function is continuous on the
closure of its domain then it is in fact smooth.
Admissible trading strategies $\pi$ are not assumed to satisfy (\ref{novikov})
and therefore the verification theorem \ref{verification} cannot be applied.

\begin{theorem} Assume that $\tilde V$ is continuous
on $[0,T]\times (0,\infty)$. Then $\tilde{V}=\hat{V}$.
\end{theorem}
\noindent {\it Proof.}\
 Since $\hat\pi\in \tilde K$ is an admissible control for the dual problem, we have
\[
0 \leq \tilde{V}(t,y)\leq E[\tilde U(\hat Y_T)]
\leq K(1+ y^{\frac{p}{p-1}}), y>0.
\]
We have $0\leq \tilde{V}(t,y),
\hat{V}(t,y)\leq K(1+ y^{\frac{p}{p-1}})$. Let $h(y)=y+y^{-m}$
with $m>\frac{p}{1-p}$. Then
\[
h(y)>0,\ h''(y)>0
\]
for $y>0$. Let
\[
w(t,y)=\frac{\tilde{V}(t,y)-\hat{V}(t,y)}{e^{\lambda t}h(y)}
\]
where $\lambda$ is a constant to be determined later.
Then $w$ is continuous on $[0,T]\times (0,\infty)$ and
\[
w(T,y)=0, \ \lim_{y\rightarrow 0}w(t,y)=0,\ \lim_{y\rightarrow
\infty}w(t,y)=0.
\]

\noindent {\bf Claim:} $w(t,y)\leq 0$ for $(t,y)\in [0,T]\times [0,\infty)$.

If not, then there is a point $(t_0,y_0)\in [0,T)\times(0,\infty)$
such that
\[
w_0:=w(t_0,y_0)=\sup_{[0,T)\times (0,\infty)}w(t,y)>0.
\]
Let
$$
\phi(t,y):=\hat{V}(t,y)+w_0e^{\lambda t}h(y)
$$
be a test function which satisfies
$\tilde{V}(t,y)\leq \phi(t,y)$ and $\tilde{V}(t_0,y_0)=\phi(t_0,y_0)$.
Since $\tilde{V}$ is a viscosity subsolution of (\ref{tHJB})
we have
\begin{equation} \label{e11}
\phi_t+ \inf_{\tilde{\pi}\in
\tilde{K}}\left\{\frac{1}{2}|\theta(t)+\sigma(t)^{-1}
\tilde{\pi}|^2y^2\phi_{yy}\right\}\geq
0
\end{equation}
at $(t_0,y_0)$.
Substituting $\phi$ into (\ref{e11}), also noting that
$\phi_{yy}>0$ and $\hat V$ is a solution of (\ref{dualHJB}),  we get
\[
w_0 e^{\lambda t}\left(
\frac{1}{2}|\hat{\theta}(t)|^2y^2h''+\lambda h\right)\geq 0
\]
at $(t_0,y_0)$. Substituting  $h(y)=y+y^{-m}$ into the above inequality, we obtain
\[
\lambda y_0^{m+1}+\lambda+\frac{1}{2}|\hat{\theta}(t)|^2m(m+1)\geq
0.
\]
This leads to a contradiction if we choose
$\lambda<-\frac{1}{2}\theta_1^2m(m+1)$ where
$\theta_1=\max_{0\leq t\leq T}|\hat\theta(t)|$.
 This proves that $\tilde{V}(t,y)\leq \hat{V}(t,y)$. Similarly,
 we can show that $\hat{V}(t,y)\leq \tilde{V}(t,y)$. \qed

\begin{lemma} Let Assumption \ref{utility} hold.
Then the value function $V(t,x)$ satisfies
$$0\leq V(t,x)\leq
\tilde{L}(1+x^p)$$
 for some constant $\tilde{L}$.
\end{lemma}
\noindent {\it Proof.}
Define a stochastic process for $0\leq t\leq T$ by
$$ H_t=\exp\left(-\int_0^t \theta(s)dW_s
-{1\over 2}\int_0^t |\theta(s)|^2 ds\right).$$ Assumption
\ref{parabolicity} and the Novikov condition imply that $H$ is a
positive martingale. Define an equivalent probability measure $Q$
by ${dQ\over dP}=H_T$. Then the Girsanov theorem implies that
$W^0_t=W_t+\int_0^t \theta(s)ds$ is a $Q$-Brownian motion
 and the wealth process $X$ is a $Q$-supermartingale
 for $0\leq t\leq T$. Therefore
$E_Q[X_T]\leq x$. Note also that
$$ {dP\over dQ}=\tilde H_T
:=\exp\left(\int_0^t \theta(s)dW^0_s -{1\over 2}\int_0^t
|\theta(s)|^2 ds\right)
$$
Let $\tilde p=1/p$ and $\tilde q=1/(1-p)$, applying the Holder inequality, we get
$$ E[X_T^p]=E_Q[X_T^p{dP\over dQ}]
\leq \left(E_Q[(X_T^p)^{\tilde p}]\right)^{1/\tilde p}
\left(E_Q[(\tilde H_T)^{\tilde q}]\right)^{1/\tilde q}
\leq x^p (E_Q[\tilde H_T^{\tilde q}])^{1/\tilde q}.
$$
Since
$$ \tilde H_T^{\tilde q}
= \exp\left(\int_0^T \tilde q\theta(s)dW^0_s - {1\over 2}\int_0^T
\tilde q^2 |\theta(s)|^2 ds\right) \exp\left({1\over 2}\int_0^T
(\tilde q^2-\tilde q)|\theta(s)|^2 ds\right)
$$
and $W^0$ is $Q$-Brownian motion, we get
$$ E_Q[\tilde H_T^{\tilde q}]
= \exp\left({1\over 2}\int_0^T (\tilde q^2-\tilde q)|\theta(s)|^2
ds\right)
$$
which results in
$$ (E_Q[\tilde H_T^{\tilde q}])^{1/\tilde q}
= \exp\left({p\over 2(1-p)} \int_0^T |\theta(s)|^2 ds\right).
$$
Putting everything together, we get from (\ref{growth}) that
$$
V(t,x)=\sup_{\pi\in A(t,x)}E[U(X_T)] \leq  L \left(1+ \sup_{\pi\in
A(t,x)}E[X^p_T] \right) \leq \tilde L(1+x^p)$$ where
$\tilde{L}=Le^{{p\over 2(1-p)} \int_0^T |\theta(s)|^2 ds}$. \qed

\begin{lemma} Assume
the value function $V$ is continuous on $[0,T]\times [0,\infty)$. Then $V$ is a viscosity solution to the HJB equation:
$$ -V_t(t,x)- H(t,x,V_x(t,x), V_{xx}(t,x))=0$$
with the terminal condition $V(T,x)=U(x)$ and the boundary condition
$V(t,0)=0$.
\end{lemma}
\noindent {\it Proof.} Since $K$ is a cone we
know that the Hamiltonian $H$ defined
in (\ref{hamiltonian}) is $\infty$ if $M>0$ and is either 0 or $\infty$ if $M=0$. Therefore, if $H$ is positive  at some point
$(t,x,p,M)$ we must have $M<0$.  Applying Lemma~\ref{L1}, we
can write
$$ H(t,x,p,M)=x^2M\inf\left\{\pi'b(t){p\over xM}+{1\over 2}|\sigma(t)'\pi|^2
\right\}
=-{p^2\over 2M}|\hat\theta(t)|^2.$$
It is clear that
$H$ is continuous at any point where it is positive. The remaining
proof that $V$ is a viscosity supersolution and a subsolution is the same as that of \cite{pham}, Prop.~4.3.1 and Prop.~4.3.2. The only difference is that we do not use the function $G$ as in the proof of Prop.~4.3.2 for subsolution property. The function
$G$ ensures that  $H$ is continuous at any point where it is positive, which has been established directly. \qed

\begin{remark}{\rm
In fact, we only need to assume that the value function $V$ is locally bounded on $[0,T]\times [0,\infty)$ to get the viscosity property. We need to define its upper-semicontinuous envelope $V^*$ and lower-semicontinuous envelope $V_*$ on $[0,T]\times [0,\infty)$ by
$$ V^*(t,x)=\limsup_{(t',x')\to (t,x)} V(t',x'),\quad
 V_*(t,x)=\liminf_{(t',x')\to (t,x)} V(t',x')
$$
and use $V^*$ (and $V_*$)
instead of $V$ in the definition of viscosity subsolution
(and supersolution). Since $V$ may be discontinuous at boundary of  $[0,T]\times [0,\infty)$ it is much subtle to define the proper terminal and boundary conditions, see \cite{pham, touzi} for details. This is the main reason we assume that $V$ is continuous on $[0,T]\times [0,\infty)$. In general, one needs to add
some strong conditions to ensure the continuity of $V$ on the closure of its domain, see \cite{fs}.
}\end{remark}

We can now state the main result of this section.
\begin{theorem} \label{smoothV}
Assume that the value function $V$ is continuous on $[0,T]\times
[0,\infty)$. Then $V=u$ and $V$ is a classical solution to the HJB
equation (\ref{HJB}).
\end{theorem}

\noindent {\it Proof.}\ Let $H=\max\{\tilde{L},\tilde{K}\}$. Then
$0\leq V(t,x), u(t,x)\leq H(1+x^p)$. Let $h(x)=(x+1)^q$ with
$p<q<1$. Then
\[
h(x)>0,\ h'(x)>0,\ h''(x)<0
\]
for $x\geq 0$. Define
\[
w(t,x)=\frac{V(t,x)-u(t,x)}{e^{\lambda t}h(x)}
\]
where $\lambda$ is a constant to be determined later.
Then $w(t,x)\in C([0,T]\times [0,\infty)$ and
$$ w(T,x)=0,\ w(t,0)=0,\ \lim_{x\rightarrow \infty}w(t,x)=0.$$

\noindent {\bf Claim:} $w(t,x)\leq 0$ for all $(t,x)\in[0,T]\times [0,\infty)$.

If not, then there is a point $(t_0,x_0)\in (0,T)\times(0,\infty)$
such that
\[
w_0:=w(t_0,x_0)=\sup_{[0,T]\times[0,\infty)}w(t,x)>0.
\]
Let
$$
\phi(t,x)=u(t,x)+w_0e^{\lambda t}h(x)
$$
be a test function which satisfies
$V(t,x)\leq \phi(t,x)$ and $V(t_0,x_0)=\phi(t_0,x_0)$.
Since $V$ is a viscosity subsolution of (\ref{HJB}) we have
\begin{equation}\label{e13}
\phi_t+ \sup_{\pi\in
K}\left\{\frac{1}{2}|\pi'\sigma(t)|^2x^2\phi_{xx}+\pi' b(t) x\phi_x\right\}\geq 0
\end{equation}
at $(t_0,x_0)$. Substituting $\phi$
into (\ref{e13}), also noting that
$u$ is a solution of (\ref{HJB}), we get
$$ w_0 e^{\lambda t}\left(\sup_{\pi\in
K}\left\{\frac{1}{2}|\pi'\sigma(t)|^2x^2h''+\pi'b(t)xh'\right\}+\lambda h\right)
\geq 0
$$
at $(t_0,x_0)$.
Applying Lemma~\ref{L1} we obtain
$$
-\frac{1}{2}|\hat{\theta}(t_0)|^2\frac{h'^2(x_0)}{h''(x_0)}+\lambda
h(x_0)\geq 0.
$$
Since $h(x)=(x+1)^q$, the above inequality becomes
\[
-\frac{1}{2}|\hat{\theta}(t_0)|^2\frac{q(x_0+1)^q}{(q-1)}+\lambda(x_0+1)^q\geq
0.
\]
This leads to a contradiction if we choose
$\lambda<-\frac{q}{2(1-q)}\theta_1^2$. This proves that $V(t,x)\leq
u(t,x)$. Similarly, we can show $u(t,x)\leq V(t,x)$. \qed

\section{Applications}
In this section we present two examples which can be solved with
the main results of the paper. The first one is the efficient
frontier of utility and CVaR and the second one is the
preservation of monotonicity of the absolute risk aversion.

\subsection{Efficient Frontier of Utility and CVaR}
In the standard utility maximization theory the  risk is not considered. However, in practice one often needs to find the optimal tradeoff between return and
risk. This is the fundamental idea of the Markowitz's mean variance
efficient frontier theory. In \cite{zheng} the problem of the efficient frontier of utility and CVaR is discussed. A utility loss random variable
$Z$ is defined by $Z=U(x_0)-U(X_T)$, which represents the risk associated with a trading strategy $\pi$ in comparison with a riskfree strategy $\pi=0$.
Two common risk measures are VaR and CVaR. Given a
number $\beta\in (0,1)$ (close to 1) the $\beta$-VaR of  $Z$ is defined by
$$
 {\rm VaR}_\beta=\min\{z: P(Z\leq z)\geq \beta\}
 $$
and the $\beta$-CVaR of $Z$ is defined by
$$ {\rm CVaR}_\beta=\mbox{mean of the $\beta$-tail distribution of $Z$}$$
where the $\beta$-tail distribution $F_\beta(z)$ is defined by
$$ F_\beta(z)=\left\{\begin{array}{ll}
0&\mbox{for } z< {\rm VaR}_\beta\\
{P(Z\leq z)-\beta\over 1-\beta}& \mbox{for } z\geq {\rm VaR}_\beta.
\end{array}\right.
$$
A fundamental minimization formula is established in
\cite{ru02}, Theorem 10, to compute ${\rm VaR}_\beta$
and ${\rm CVaR}_\beta$  by solving  a convex minimization
problem in which the minimum value is ${\rm CVaR}_\beta$
and the left end point
of the minimum solution set gives ${\rm VaR}_\beta$. Specifically,
$$
{\rm CVaR}_\beta = \min_{y} [ y+ \delta E(Z-y)^+]
$$
where $\delta=(1-\beta)^{-1}$.
If $y^*$ is the left endpoint of the minimum solution set, then
${\rm VaR}_\beta=y^*$.

The following optimization problem is discussed in \cite{zheng}:
$$
\sup_{\pi} (E[U(X_T)] - \lambda {\rm CVaR}_\beta)
\;\mbox{ subject to (\ref{wealth})}
$$
where $\lambda$ is a nonnegative parameter.
 $\lambda=0$ corresponds to the utility maximization while
 $\lambda\to\infty$ to the CVaR minimization. The  efficient frontier
of  utility and CVaR can be determined by first solving
a parametric utility maximization problem
\begin{equation} \label{parametric}
u(x_0,y) =  \sup_{\pi} E [U^y(X_T)]
\;\mbox{ subject to (\ref{wealth})}, \label{value_y}
\end{equation}
where
$$
U^y(x)= U(x) - \lambda \delta(U(x_0)-U(x) -y)^+ + \lambda \delta (U(x_0)-y)^+,
$$
and then solving a scalar concave maximization  problem
\begin{equation} \label{2stage}
u(x_0)=\sup_{y} (u(x_0,y)-\lambda\delta(U(x_0)-y)^+-\lambda y).
\end{equation}

There exists an optimal solution to the first stage problem (\ref{value_y})
for every fixed $y$ under some additional conditions on 
$U$ (strictly increasing, strictly concave, $C^1$, and $U'(0)=\infty$), and
there exists an opitmal solution to the second stage problem (\ref{2stage})
as the objective function is concave, Lipschitz continuous, and tends to
$-\infty$ as $y$ tends to $\infty$, see \cite{zheng} for details. 

Note that if $U$ satisfies Assumption  \ref{utility} then so does $U^y$ for
every fixed $y$. Therefore Theorems \ref{maintheorem} and \ref{verification}
hold true for parametric utility maximization problem (\ref{parametric}).
In particular, we can construct a smooth classical solution to HJB equation 
(\ref{HJB}) with $U$ being replaced by $U^y$ and 
show that the value function is equal to that smooth solution 
if admissible trading strategies satisfy the integrability condition (\ref{novikov})
for every fixed $y$. This opens the way to solve the first stage problem
with the standard numerical method for nonlinear PDEs and to find the
parametric optimal control and optimal value for problem (\ref{parametric}).

\subsection{Monotonicity of Absolute Risk Aversion Measure}
In this subsection we assume $U\in C^2$. The Arrow-Pratt measure of absolute risk aversion for a utility function
$U$ is defined by
$$  R(x)= -{U''(x)\over U'(x)}.$$
$R$ is a constant for exponential utility functions  and is a decreasing
function for power and logarithmic utility functions. Since $R(x)=-(\ln U'(x))'$
it is clear that $R$ is increasing (decreasing) if and only if $\ln U'(x)$
is concave (convex). For the value function $V(t,x)$ with $V(T,x)=U(x)$ we
may define a dynamic Arrow-Pratt measure of absolute risk version by
$$ R(t,x)=-{V_{xx}(t,x)\over V_x(t,x)}$$
provided all derivatives are well defined.
The monotonicity properties of
optimal investment strategies is discussed in \cite{Bore07}
which shows that  that $R(t,x)$ inherits the
monotonicity of $R(x)$ with the martingale approach. Here we give a
new proof with the PDE approach and the duality method.
We also extend the results of \cite{Bore07} as we do not need the Inada
condition. The next result is needed in proving the monotonicity of $R(t,x)$.

\begin{lemma} \label{RA}
 Suppose that
$\tilde{U}\in C^1$ and $y\tilde{U}'(y)-\tilde{U}(y)$ is
convex (concave) in $y$. Then $y\hat{V}_y(t,y)-\hat{V}(t,y)$ is strictly
convex (concave) in $y$ for $t<T$. 
\end{lemma}

\noindent{\it Proof.}\ Let $w(t,y)=y\hat{V}_y(t,y)-\hat{V}(t,y)$.
From (\ref{hatV}), we get
\[
w(t,y)=\frac{1}{2\sqrt{\pi\tau}} e^{-\frac{1}{4}\tau}
 \int_0^\infty
e^{-\frac{(\ln \xi)^2}{4\tau}} \xi^{-\frac{3}{2}}
[(y\xi)\tilde{U}'(y\xi)-\tilde{U}(y\xi)]d\xi.
\]
This implies the convexity of $w(t,y)$ in $y$ for $t<T$. A simple computation
yields the following equation
\[
w_t+\frac{1}{2}|\hat{\theta}(t)|^2 y^2w_{yy} =0,\ y>0,\;0\leq t<T.
\]
As in Lemma \ref{LemhatV}, we deduce that if $w_{yy}(t_0,y_0)=0$
for some $(t_0,y_0)$ with $t_0<T$ then $\tilde{U}(y)=C_1 y\ln
y+C_2 y+C_3$ with constants $C_1, C_2, C_3$, which contradicts 
the assumption \ref{utility}. Therefore $w(t,y)$ is strictly convex in $y$ for $t<T$.
\qed

The next theorem shows that $\ln V_x(t,x)$ preserves the
convexity (concavity) of $\ln U'(x)$.

\begin{theorem} \label{RA2}
Let the assumptions of Theorem \ref{maintheorem} hold and
let $U\in C^1$ and $U$ be strictly increasing and strictly concave.
Then $\ln V_x(t,x)$ is strictly convex (concave) for $t<T$ if $\ln
U'(x)$ is convex (concave).
\end{theorem}

\noindent {\it Proof.} Assume $\ln U'(x)$ is concave. As in the
proof of Lemma 4.1 in \cite{Bore07}, we conclude that
$\tilde{U}'(e^z)$ is convex. Then $y\tilde{U}'(y)-\tilde{U}(y)$ is
convex. From Lemma \ref{RA}, we see that
$y\hat{V}_y(t,y)-\hat{V}(t,y)$ is strictly convex for $t<T$, i.e.,
$(y\hat{V}_y(t,y)-\hat{V}(t,y))_{yy}>0$ for $t<T$.
A direct computation implies
\[
(y\hat{V}_y(t,y)-\hat{V}(t,y))_{yy}=(y\hat{V}_{yy})_y=\frac{(\ln
V_x)_{xx}}{V_{xx}((\ln V_x)_{x})^2}.
\]
Since $V$ is strictly concave in $x$ for $t\in [0,T)$,
we conclude that $(\ln V_x)_{xx}<0$ and $\ln V_x(t,x)$ is strictly
concave for $t<T$. \qed

\begin{corollary} \label{RA3}
In addition to the assumptions of Theorem \ref{RA2}, assume that
$U\in C^2$. Then $R(t,\cdot)$ inherits the monotonicity of $R$
for $t<T$.
Furthermore, $R(t,\cdot)$ is strictly increasing (decreasing) for $t<T$
if $R$ is increasing (decreasing).
\end{corollary}

\bigskip\noindent
{\bf Acknowledgement}. The authors thank Martin Schweizer and Nizar Touzi for the useful discussions and comments on the contents of the paper.


\begin{thebibliography}{99}

\bibitem{BG08} B. Bian and P. Guan, Convexity preserving for fully nonlinear parabolic integro-differential equations,
Methods Appl.~Anal., 15 (2008), pp.~39-52.

\bibitem{BG09} B. Bian and P. Guan, A microscopic convexity principle for nonlinear partial differential Eequations, Invent.~Math., 177 (2009), pp.~307-335.

\bibitem{Bore07} C. Borell,  Monotonicity properties of
optimal investment strategies for log-Brownian asset prices, 
Math. Finance, 17 (2007), pp.~143-153.

\bibitem{btz} B. Bouchard, N. Touzi, and A. Zeghal,
Dual formulation of the utility maximization problem: the case of nonsmooth utility, Ann.~Appl.~Probab., 14 (2004), pp.~678-717.



\bibitem{dpt} G. Deelstra, H. Pham, and N. Touzi, Dual formulation of the utility maximization problem under transaction costs, Ann.~Appl.~Probab.,
 11 (2001), pp.~1353-1383.


\bibitem{fs} W. Fleming and M. Soner, 
Controlled Markov Processes and Viscosity  Solutions, Springer, 1993.


\bibitem{Fri} A. Friedman, Partial Differential Equations of Parabolic Type,
Prentice-Hall, 1964.

\bibitem{JT04} S. Janson and J. Tysk, Preservation of convexity of solutions to parabolic equations, J. Differential Equations, 206 (2004), pp.~182-226.

\bibitem{ks98} I. Karatzas and S.E. Shreve,
Methods of Mathematical Finance, Springer, 1998.

\bibitem{ks} D. Kramkov and W. Schachermayer, The asymptotic elasticity of utility functions and optimal investment in incomplete markets,
Ann.~Appl.~Probab.,  9 (1999), pp.~904-950.

\bibitem{LM} P. Lions and M. Musiela, Convexity of solutions of parabolic equations,  C. R. Math.~Acad.~Sci.~Paris, 342 (2006), pp.~915-921.

\bibitem{pham} H. Pham, Continuous-time Stochastic Control and Optimization with Financial Applications, Springer, 2009.

\bibitem{ru02} R.T. Rockafellar and S. Uryasev, Conditional value-at-risk for general loss distributions, J. Banking Finance, 26 (2002), pp.~1443-1471.

\bibitem{ss} S. Shreve and M. Soner, Optimal investment and consumption with transaction costs. Ann.~Appl.~Probab.,  4 (1994), pp.~609-692.

\bibitem{touzi} N. Touzi,  Stochastic Control Problems, Viscosity Solutions,
and Application to Finance, Scuola Normale Superiore, 2002.

\bibitem{wz09} N. Westray and H. Zheng, Constrained nonsmooth utility maximization without quadratic inf-convolution,  Stochastic Process.~Appl., 
119 (2009), pp.~1561-1579.

\bibitem{wz10} N. Westray and H. Zheng, Minimal sufficient conditions for a primal optimizer in nonsmooth utility maximization,
Finance Stoch., forthcoming, 2010.


\bibitem{XS2} G. Xu and S. Shreve, A duality method for optimal and investment under short-selling prohibition, II. constant market
 coefficients, Ann.~Appl.~Probab.,
 2 (1992), pp.~314-328.

\bibitem{zheng} H. Zheng, Efficient frontier of utility and CVaR,
Math.~Methods Oper.~Res., 70 (2009), pp.~129-148.


\end{thebibliography}
\end{document}